\documentclass[aps,prl,floatfix,twocolumn,nopacs,superscriptaddress]{revtex4-1}

\usepackage[usenames,dvipsnames]{color}

\newcommand{\vect}[1]{\boldsymbol{#1}}

\usepackage{graphicx}
\usepackage{epstopdf}
\usepackage{amsfonts,amssymb,amsmath}
\usepackage{bm}
\usepackage{hyperref}

\begin{document}
\title{
Topological quantum phase transition in magnetic topological insulator 
upon magnetization rotation
}

\author{Minoru~Kawamura}
\email{minoru@riken.jp}
\affiliation{
	RIKEN Center for Emergent Matter Science, Wako  351-0198, Japan\\
}

\author{Masataka~Mogi}
\affiliation{
	Department of Applied Physics	and Quantum-Phase Electronics Center (QPEC), University of Tokyo,  Tokyo 113-8656, Japan\\
}

\author{Ryutaro~Yoshimi}
\affiliation{
	RIKEN Center for Emergent Matter Science, Wako  351-0198, Japan\\
}

\author{Atsushi~Tsukazaki}
\affiliation{
	Institute for Materials Research, Tohoku University, Sendai 980-8577, Japan\\
}

\author{Yusuke~Kozuka}
\affiliation{
	Department of Applied Physics and Quantum-Phase Electronics Center (QPEC), University of Tokyo,  Tokyo 113-8656, Japan\\
}

\author{Kei.~S.~Takahashi}
\affiliation{
	RIKEN Center for Emergent Matter Science, Wako  351-0198, Japan\\
}
\affiliation{
	PRESTO, Japan Science and Technology Agency (JST),  Tokyo 102-0075, Japan
}

\author{Masashi~Kawasaki}
\affiliation{
	RIKEN Center for Emergent Matter Science, Wako  351-0198, Japan\\
}
\affiliation{
	Department of Applied Physics	and Quantum-Phase Electronics Center (QPEC), University of Tokyo,  Tokyo 113-8656, Japan\\
}

\author{Yoshinori~Tokura}
\affiliation{
	RIKEN Center for Emergent Matter Science, Wako  351-0198, Japan\\
}
\affiliation{
	Department of Applied Physics	and Quantum-Phase Electronics Center (QPEC), University of Tokyo,  Tokyo 113-8656, Japan\\
}

\date{\today}

\begin{abstract}
We report a continuous phase transition between quantum-anomalous-Hall and trivial-insulator phases  in a magnetic topological insulator upon magnetization rotation.
The Hall conductivity transits from one plateau of quantized Hall conductivity $e^2/h$ to the other plateau of zero Hall conductivity.
The transition curves taken at various temperatures cross almost at a single point, exemplifying the critical behavior of the transition.
The slope of the transition curves follows a power-law temperature dependence with a critical exponent of $-0.61$.
This suggests a common underlying origin in the plateau transitions between the QAH and quantum Hall systems, which is a percolation of  one-dimensional chiral edge channels.
\end{abstract}

\maketitle

A topological phase of matter may undergo a quantum phase transition accompanied by a change in a topological invariant number when material compositions or the sample dimensions are changed\cite{Bernevig2006:science, Konig2007:science, Murakami2007:njp, Zhang2010:natphys, Wang2014:prb, Wang2015:physscr, Kou2015:natcommun, Feng2015:prl, Mogi2017:natmater, Mogi2017:sciadv, Chang2016:prl} .
Such transitions termed topological quantum phase transitions have been one of the central issues of discussion in the study of topological phases of matter\cite{Hasan2010:rmp, Qi2011:rmp, Tokura2017:natphys}.

Three-dimensional topological insulator (3DTI) originating from strong spin-orbit interaction is one of the most distinguished topological phases of matter, which consists of an insulating bulk and a gapless  surface state with a linear dispersion relation\cite{Fu2007:prl}.
When magnetic atoms are introduced to a 3DTI and it becomes a ferromagnet, an energy gap opens in the dispersion relation on the surfaces perpendicular to the magnetization due to the exchange interaction.
On the other hand, the side surfaces parallel to the magnetization remain gapless, forming a one-dimensional chiral edge channel.
As a result, the  Hall resistance is quantized to $h/e^2$ and the longitudinal resistance vanishes\cite{Yu2010:science}.
This phenomenon known as quantum anomalous Hall (QAH) effect reflects the non-trivial topology (Chern number $C$ = 1) of the surface state of a magnetic 3DTI. 
The QAH effect has been observed experimentally in thin films of
Cr-\cite{Chang2013:science, Checkelsky2014:natphys, Mogi2015:apl} or V-doped\cite{Chang2015:natmater, Grauer2017:prl} (Bi, Sb)$_2$Te$_3$
grown by molecular beam epitaxy.
Besides that, the surface state of a 3DTI in a thin film form can be gapped  when the opposite surfaces hybridize via quantum tunneling\cite{Zhang2010:natphys, Wang2014:prb, Linder2009:prb}.
In this case, the surface state becomes a trivial insulator ($C$ = 0) without chiral edge channels.
Therefore a topological quantum phase transition between the QAH ($C$ = 1) and trivial-insulator ($C$ = 0) phases is naturally anticipated to occur with varying the exchange interaction energy  $\Delta_{\rm ex}$.

Theoretical works by Wang {\it et al.}\cite{Wang2014:prb, Wang2015:physscr}
have addressed this issue and predict that the phase transition occurs when  $\Delta_{\rm ex}$ matches with the hybridization energy $\Delta_{\rm hy}$.
A power-law divergent behavior of the localization length is predicted as the critical point is approached.
Because the energy gap closes at the critical point, the one-dimensional chiral edge channels propagating on the side surfaces penetrate into the top and bottom surfaces.
This situation is similar to the plateau-plateau transitions of the quantum Hall (QH) effect in high magnetic fields\cite{Wei1988:prl, Koch1992:prb,  Hilke1998:nature,  Huckenstein1995:rmp, Sondhi1997:rmp, Chalker1998:jphys}.
Generally, a phase transition can be characterized by its critical behavior which is governed by the dimensionality and/or the fundamental symmetry of the system regardless of the details.
Therefore both QAH and QH systems are expected to possess the same critical exponent for the diverging localization length.

\begin{figure}[t]
	\includegraphics[width=.9\columnwidth]{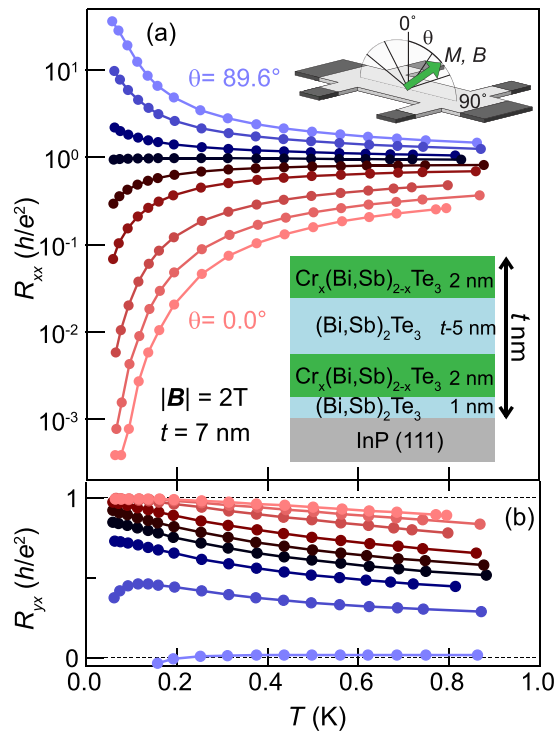}
	\caption{	
			Temperature dependence of $R_{xx}$ (a)
			and $R_{yx}$ (b) at $|\vect{B}|$ = 2 T
			under various magnetization angles $\theta$.
			The data for $\theta$ = 0.0$^\circ$, 28.9$^\circ$, 43.3$^\circ$, 56.5$^\circ$,
			 62.2$^\circ$, 66.9$^\circ$, 70.7$^\circ$, 77.8$^\circ$, and 89.6$^\circ$
			are shown from bottom to top in (a).
			$\theta$ was measured by using a Hall sensor
			from the direction normal to the film plane as shown
			in the upper inset in (a).
			The lower inset in (a) shows a schematic film structure.
			}
	\label{resistance}
\end{figure}

In this Letter, we report a critical behavior of the topological quantum phase transition between
the quantum anomalous Hall and trivial-insulator phases.
To access the critical regime of the transition experimentally, a continuous and in situ control of $\Delta_{\rm ex}$ is a key.
We choose the magnetization angle as a knob to control $\Delta_{\rm ex}$ following the earlier works\cite{Kou2015:natcommun, Mogi2017:natmater}.
Because, in a thin film form of a magnetic 3DTI,  only the magnetization component perpendicular to the film plane contributes to the energy gap formation in the QAH phase, we can control $\Delta_{\rm ex}$ by rotating the magnetization with respect to the film plane.
We observed the Hall conductivity  $\sigma_{xy}$ transits from one plateau of quantized Hall conductivity $e^2/h$ to the other plateau of zero Hall conductivity.
The transition curves of $\sigma_{xy}$ obtained under various temperatures crossed at a single point and the critical exponent of the phase transition was evaluated as $\kappa$ = $-0.61$ from the slope of $\sigma_{xy}$.
Although the phase transition driven by magnetization rotation was demonstrated in the earlier works\cite{Kou2015:natcommun,  Mogi2017:natmater}, the critical regime of the transition, including the critical exponent, has not been explored to date.

Experiments were conducted using heterostructure films of  (Bi, Sb)$_2$Te$_3$ sandwiched by 2-nm-thick Cr-doped (Bi, Sb)$_2$Te$_3$ layers [lower inset in Fig.~\ref{resistance}(a)] following the previous work\cite{Mogi2017:natmater}.
Cr (12\%) was doped selectively to 2-nm-thick layers in the vicinity of the top and bottom surfaces
in $t$-nm-thick sample, so that the conduction electrons on the top (bottom) surface interact with the upper (lower) Cr-doped layer through the exchange interaction.
It is known that the observable temperature of the QAH effect increases substantially in the heterostructure films compared to the homogeneously Cr-doped films\cite{Mogi2015:apl}. 
The Cr concentration and the structure near the surfaces are common
to all the films to keep the exchange interaction the same
while middle pristine layer thickness was varied.
Transport measurements were conducted at low temperatures by using a dilution refrigerator equipped a superconducting solenoid and a single-axis sample rotator.
A standard lock-in technique was employed for the resistance measurement with low excitation currents (1 nA or 10 nA) at low frequencies (3 Hz or 0.3 Hz).
The QAH effect with quantized Hall resistance and vanishing longitudinal resistance is clearly observed in all the samples at temperature $T$ = 60 mK (Supplemental Material \cite{supple}).

When the magnetization is rotated away from the direction normal to the top/bottom surfaces, the temperature dependence of $R_{xx}$ demonstrates a remarkable metal-insulator transition behavior as shown in Fig.~\ref{resistance}(a).
To rotate the magnetization, the direction of $\vect{B}$ is changed with respect to the film plane.
In the present work, we applied  $|\vect{B}|$ = 2 T so that the magnetization follows the direction of the external magnetic field overcoming the perpendicular magnetic anisotropy of Cr-doped  (Bi, Sb)$_2$Te$_3$\cite{Chang2013:science}.
When $\theta$ = 0$^\circ$ [$\vect{B}$ perpendicular to the film plane, pink in Figs. \ref{resistance}(a)  and \ref{resistance}(b)], $R_{xx}$  goes to zero and  $R_{yx}$  tends to be saturated to $h/e^2$ with decreasing $T$.
This $T$ dependence shows that the surface state is in the QAH phase.
When $\theta$ =  89.6$^\circ$ [$\vect{B}$ parallel to the film plane, blue in Figs. \ref{resistance}(a)  and \ref{resistance}(b)],  $R_{xx}$ diverges and $R_{yx}$ becomes nearly zero as $T$ is lowered, indicating that the surface state is in the trivial-insulator phase.
The transition between the QAH and the trivial-insulator phases occurs at around $\theta$ = 66.9$^\circ$,
where $R_{xx}$ is nearly temperature independent.

\begin{figure*}[t]
	\includegraphics[width=2.\columnwidth]{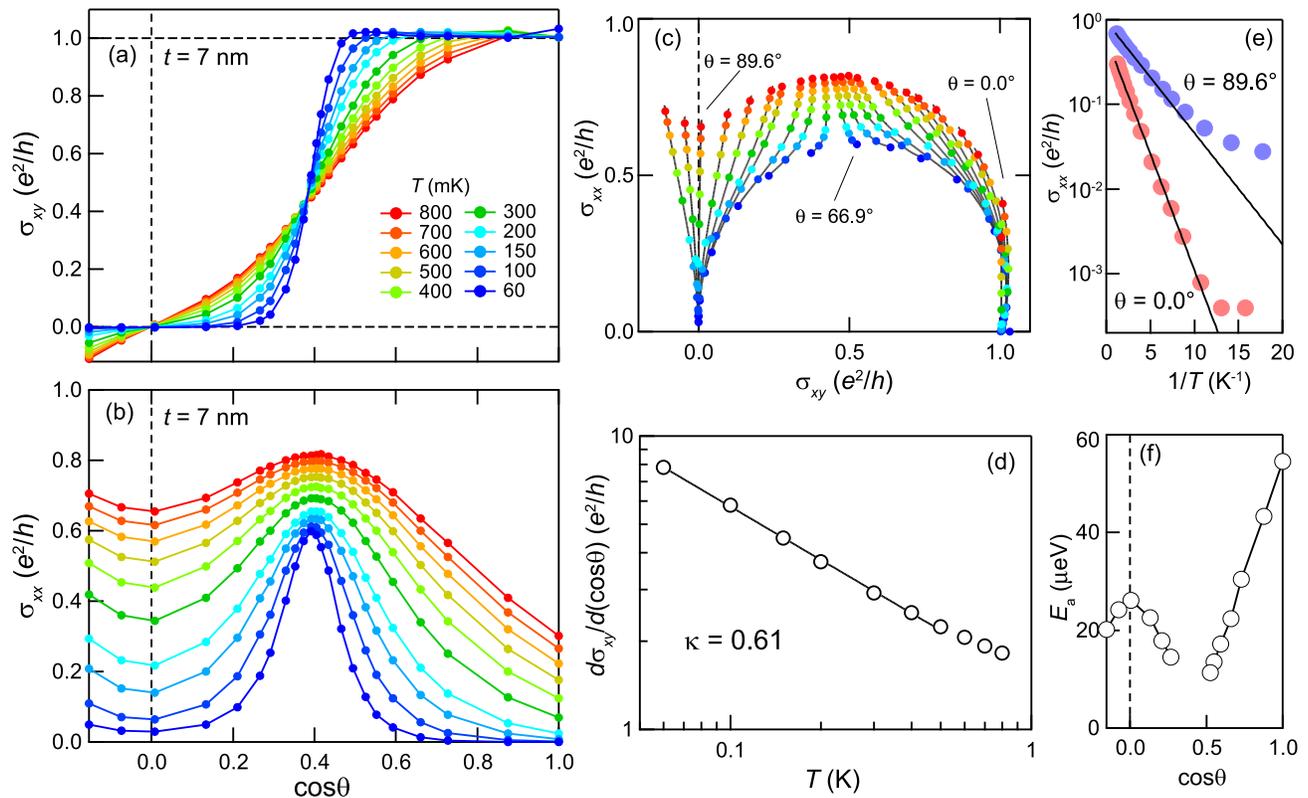}
	\caption{
			Magnetization angle dependence of $\sigma_{xy}$ (a)
			and $\sigma_{xx}$ (b) for $t$ = 7 nm sample plotted as a function
			of $\cos\theta$ under various temperatures
			ranging from 60 mK to 800 mK. 
			To make these plots, we first interpolated the $T$ dependence
			of $R_{xx}$ and $R_{yx}$ for each $\theta$ and calculated
			the values of $\sigma_{xx}(\theta,T)$ and $\sigma_{xy}(\theta, T)$
			using the tensor relation
			$\sigma_{xx} = \rho_{xx}/(\rho_{xx}^2 + \rho_{yx}^2)$ and 
			$\sigma_{xy} = \rho_{yx}/(\rho_{xx}^2 + \rho_{yx}^2)$.
			Then the conductivity data at the representative temperatures
			are plotted as a function of $\cos\theta$.
			(c)
			The conductivity data for various field angles
			and temperatures plotted on the $\sigma_{xx}$-$\sigma_{xy}$ plane.
			Each black curve connects the data points
			belonging to the same field angle.
			(d)
			The slope of the $\sigma_{xy}$ transition
			[$d\sigma_{xy}/d (\cos\theta )$] plotted
			as a function of temperature.
			The data are fitted to  $d\sigma_{xy}/d (\cos\theta ) = A T^{-\kappa}$ 
			with $\kappa = 0.61 \pm 0.01$ (solid line).
			(e) Log-scale plot of $\sigma_{xx}$ is shown as a function of $1/T$
			for $\theta$ = 0.0$^\circ$ (pink) and 89.6$^\circ$ (blue). 			
			The fitting results are shown by black lines.
			(f)
			Variation of the activation energy $E_{a}$
			 as a function of  $\cos\theta$. 
			}
	\label{conductivity}
\end{figure*}

Topological aspects of the phase transition can be seen more clearly in the $\theta$ dependence of the Hall conductivity [Fig.~\ref{conductivity}(a)] which is a measure of the Berry curvature in the momentum space\cite{Thouless1982:prl}.
The observed transition of $\sigma_{xy}$ from $e^2/h$ to zero corresponds to  the change in the Berry curvature from $\pi$ to zero.
On the right hand side of the transition, $\sigma_{xy}$ = $e^2/h$ demonstrates the existence of a one-dimensional chiral edge channel, while, on the left hand side of the transition, $\sigma_{xy}$ = 0 indicates the absence of the edge channel.
The transition in $\sigma_{xy}$ is  accompanied by a peak in $\sigma_{xx}$ [Fig.~\ref{conductivity}(b)]. 
As $T$ is increased,  the transition slope in $\sigma_{xy}$ becomes gentle and the $\sigma_{xx}$ peak becomes wide.
The  $\sigma_{xy}$-$\cos\theta$ curves under various $T$ cross almost at a single point, showing a critical behavior of the transition.
The transition resembles the QH plateau-plateau transitions
observed in high magnetic fields\cite{Wei1988:prl, Koch1992:prb, Hilke1998:nature, Huckenstein1995:rmp, Sondhi1997:rmp}.
A remarkable difference is the parameter which drives the phase transition.
The transition occurs as a function of $\cos\theta$ in the present QAH system,
while it is driven by the Landau level filling factor in the QH plateau-plateau transitions.

The connection of the present  QAH system to the QH system 
is further highlighted by a semi-circle plot in Fig.~\ref{conductivity}(c), in which the data shown in Figs.~\ref{conductivity}(a) and \ref{conductivity}(b) are plotted on the $\sigma_{xy}$-$\sigma_{xx}$ plane.
The data belonging to each $\theta$ are connected by a black line.
As $T$ is decreased, the flow of the conductivity data ($\sigma_{xy}(T)$, $\sigma_{xx}(T)$)
converges to  either (0,0) or ($e^2/h$, 0)
with an unstable point at around (0.5 $e^2/h$, 0.5 $e^2/h$).
This behavior is similar to the features of the theoretically calculated renormalization group flow of conductivity tensor components in the QH system\cite{Pruisken1988:prl}.
Although the similarity is already pointed out in the earlier work\cite{Checkelsky2014:natphys} where the transition is driven by gate voltage,
Fig.~\ref{conductivity}(c) demonstrates
that the same conductivity flow is reproduced in the transition driven by magnetization rotation.

To discuss the critical behavior of the transition more quantitatively, we focus on the critical exponents. 
We analyzed the slope of $\sigma_{xy}$ [$d \sigma_{xy} / d (\cos\theta)$] at the transition point which was evaluated by a linear fitting for each temperature.
The slope plotted in log-scale decreases almost linearly with  $\log T$ as shown in Fig.~\ref{conductivity}(d).
Fitting of the data to the form $d \sigma_{xy} / d (\cos\theta) = AT^{-\kappa}$ yields the exponent $\kappa$ = 0.61$\pm$ 0.01.
The exponent is consistent with those reported in the experiments of  the QH plateau-plateau transitions, which ranges from 0.4 to 0.7\cite{Gudina2017:arxiv}.
The agreement of $\kappa$ values strongly supports that the phase transition between the  QAH  and trivial-insulator phases belongs to the same category as the QH plateau-plateau transitions.
It suggests that the Chalker-Coddinton network model\cite{Chalker1998:jphys}, which describes the transition as a percolation of the one-dimensional chiral edge channel, can be applied to the phase transition between the  QAH  and trivial-insulator phases.
The experimentally obtained critical exponent $\kappa$ is a combination of two critical exponents, namely $\kappa = 1/z\nu$, where $\nu$ and $z$ are the exponents for the localization length and phase coherence length, respectively.
Therefore, for the direct comparison with the theory,  we need an independent measurement of $z$ or $\nu$, which is future work.

Apart from the critical regime,  $\sigma_{xx}$ follows the Arrhenius-type $T$ dependence in both QAH ($\theta$ = 0.0$^\circ$)  and trivial insulator ($\theta$ = 89.6$^\circ$) phases as shown in Fig.~\ref{conductivity}(e).
Fitting analysis to the form of $\sigma_{xx} \propto \exp(-E_a/k_{B} T)$ yields the thermal activation energy
$E_a$ = 55 $\mu$eV and 26 $\mu$eV for $\theta$ = 0.0$^\circ$ and 89.6$^\circ$, respectively.
At the intermediate angles, the value of $E_{a}$ decreases with decreasing $\cos\theta$,  and starts to increase again as $\cos\theta$ further decreases [Fig.~\ref{conductivity}(f)].
This dependence appears consistent with the theoretically calculated $\Delta_{\rm ex}$ dependence of the energy gap\cite{Wang2014:prb} although the size of the energy gap
thus estimated is rather small.
Near the critical point, the temperature dependence of $\sigma_{xx}$ becomes weak as seen in Fig.~\ref{conductivity}(b)  and starts to deviate from the Arrhenius-type $T$ dependence.


\begin{figure}[t]
	\includegraphics[width=1.\columnwidth]{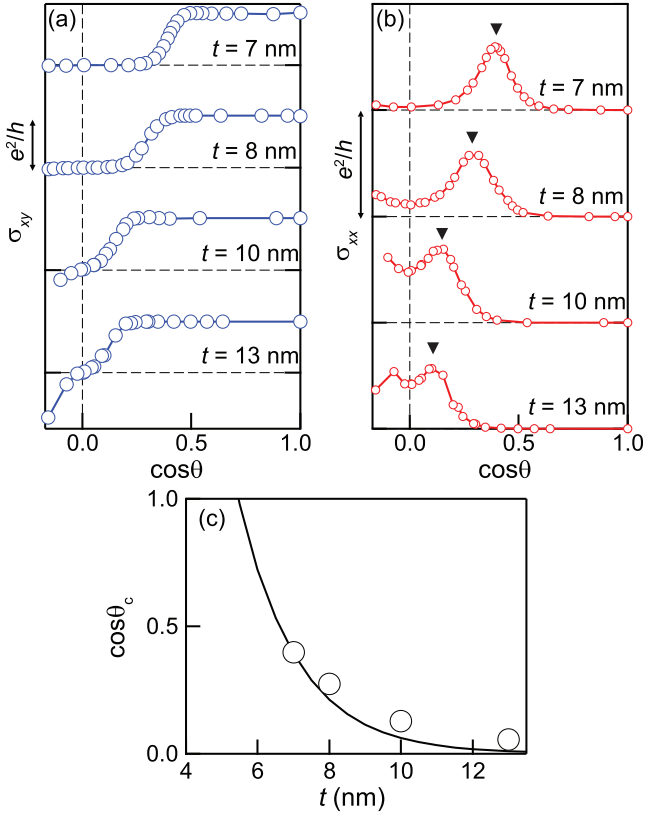}
	\caption{
	Field angle dependence of  $\sigma_{xy}$ (a) 
	and $\sigma_{xx}$ (b) at $T$ = 60 mK
	for samples with various film thicknesses $t$ = 7, 8, 10, and 13 nm.
	The data are offset for clarity. 
	Zeros are indicated by dashed lines.
	The transition angles are marked by triangles in (b).
	(c) Film thickness dependence of the transition angle determined
	from the $\sigma_{xx}$ peaks in (b).
	The solid curve shows $\cos\theta_{c} = \Delta_{\rm hy}(t)/\Delta_{\rm ex}^0$.
	}
	\label{thickness}
\end{figure}

Finally, we turn our eyes to the critical angle $\theta_{\rm c}$ where the transition takes place.
Assuming that only the magnetization component perpendicular to the film contribute to the energy gap formation, $\theta_{c}$ is given by $\Delta_{\rm ex}^0\cos\theta_{c} =  \Delta_{\rm hy}$, where $\Delta_{\rm ex}^0$ is the exchange interaction energy when the magnetization is perpendicular to the film plane.
Because $\Delta_{\rm hy}$ is smaller in the thicker film\cite{Zhang2010:natphys, Linder2009:prb}, $\cos\theta_{c}$ is expected to decrease  with increasing the film thickness $t$.
Such a trend is clearly seen  in Figs.~\ref{thickness}(a) and \ref{thickness}(b) where the $\cos\theta$ dependence of $\sigma_{xx}$ and $\sigma_{xy}$ are respectively shown for the samples with various thicknesses $t$ = 7, 8, 10, and 13 nm.
Note that the total film thickness was changed while keeping the structure of the Cr-doped layers near the surfaces to make $\Delta_{\rm ex}^0$ common to all the films [lower inset in Fig.~\ref{resistance}(a)].
The experimentally observed $t$ dependence of $\cos\theta_{c}$ is roughly explained by the ratio $\Delta_{\rm hy}(t)/\Delta_{\rm ex}^0$ as shown in Fig.~\ref{thickness}(c).
The solid curve in Fig.~\ref{thickness}(c) is drawn using the reported film thickness dependence of $\Delta_{\rm hy}(t)$ measured by photo-emission spectroscopy in (Bi, Sb)$_2$Te$_3$ thin films\cite{Zhang2010:natphys}.
We fitted their data for $t <$ 6 nm to the form of $\Delta_{\rm hy}(t) \propto \exp(-t/t_0)$ and used its extrapolation.
We also used the reported value of $\Delta_{\rm ex}^0$ =  30 meV obtained from a bulk sample of  (Bi, Sb)$_2$Te$_3$ with 10 \% of Cr doping\cite{Lee2015:pnas}, which has the similar Cr concentration to our films.
Furthermore, as seen in Fig.~\ref{thickness}(b), the value of $\sigma_{xx}$ stays finite even at $\cos\theta = 0$ in the 10-nm- and 13-nm-thick films,
showing that the surface states of the magnetic 3DTI remain conductive in these films.
These observations indicate that the trivial-insulator phase in the present study originates from the hybridization between the top and bottom surfaces,
being distinguished from the recently reported axion insulator phase of magnetic 3DTI
which originates from anti-parallel magnetization configuration of the two magnetic layers\cite{Mogi2017:natmater, Mogi2017:sciadv}.

In addition, occurrence of the phase transition itself suggests anisotropy in the exchange interaction.
In the above discussion, effects of the in-plane magnetization component are not taken into account.
The in-plane component, if it shows the exchange coupling with the surface conduction electron as the vertical component, shifts the dispersion relation in the momentum space.
The direction of the shifts are opposite to each other on the top and bottom surfaces.
Because of the opposite shifts, the energy gap of the QAH phase would not close except for $\theta = 90^\circ$ if the exchange interaction were fully isotropic (Supplemental Material \cite{supple}).
Nevertheless, the phase transition upon magnetization rotation is possible if the exchange interaction is anisotropic.
The facts that the transition is observed as a function of $\theta$ and that the observed values of $\cos\theta_{c}$ follow $\Delta_{\rm hy}(t)/\Delta_{\rm ex}^0$ suggest highly anisotropic exchange interaction in the Cr-doped (Bi, Sb)$_2$Te$_3$ film\cite{Wakatsuki2015:scirep}.

The quantitative agreement in Fig.~\ref{thickness}(c) 
also suggests that the transition angle is  less sensitive to the disorder in contrast to the activation energy.
Because the temperature dependence of $\sigma_{xx}$ is dominated by the smallest energy gap opening through from one end of the sample to the other, the activation energy will be largely reduced by disorder.
On the other hand, disorder mainly contributes to broaden the transition peak but the transition peak may stay at the same angle.
The transition width can be sensitive to the disorder.
It can be influenced by the fluctuations of $\Delta_{\rm ex}$ and $\Delta_{\rm hy}$ as well, which may arise from spatial distributions of Cr and film thickness, respectively.
The large gap between the activation energy in Fig.~\ref{conductivity}(f) and the energy scale to explain the transition angle may be attributed to the ways how the disorder in the samples affects on the measured quantities.

To conclude, the present magneto-transport study establishes a way to approach the quantum critical point of the topological quantum phase transition between the QAH and trivial-insulator phases.
We measured the critical exponent of the transition which quantitatively supports that the transition belongs to the same category as the QH plateau-plateau transition despite the different microscopic origins.
Furthermore, the occurrence of  the magnetization-rotation-driven phase transition itself suggests highly anisotropic nature of the exchange interaction.
We believe that the present work paves a way for exploring the possible quantum critical phenomena among the expanding list of topological materials.

We acknowledge fruitful discussions with Akira Furusaki, Naoto Nagaosa, Kenji Yasuda. 
This work was supported by JSPS/MEXT Grant-in-Aid for Scientific Research
(No. 15H05853, 15H05867, 17J03179, 18H04229, and 18H01155), and CREST(JPMJCR16F1), JST.

\end{document}